\documentclass[showpacs,twocolumn]{revtex4-1}
\usepackage[english]{babel}
\usepackage{graphicx}
\usepackage{amsmath, amssymb}
\usepackage{dcolumn}
\usepackage{subfigure}
\usepackage{braket}
\usepackage{amssymb}
\usepackage{amsmath}

\newcommand{\be}{\begin{equation}} 
\newcommand{\ee}{\end{equation}}
\newcommand{\bq}{\begin{eqnarray}}
\newcommand{\eq}{\end{eqnarray}}
\newcommand{\ba}{\begin{align}}
\newcommand{\ea}{\end{align}}

\begin{document}

\title{\bf Entanglement of self interacting scalar fields in an expanding spacetime}

\author{Helder Alexander $^a$}\email[]{helder@fisica.ufmg.br}
\author{Gustavo de Souza $^b$}\email[]{gdesouza@iceb.ufop.br}
\author{Paul Mansfield $^c$}\email[]{p.r.w.mansfield@durham.ac.uk}
\author{I. G. da Paz $^d$}\email[]{irismarpaz@ufpi.edu.br}
\author{Marcos Sampaio $^a$}\email[]{msampaio@fisica.ufmg.br}

\affiliation{$^a$ Universidade Federal de Minas Gerais -- Departamento de F\'{\i}sica -- ICEx \\ P.O. BOX 702,
30.161-970, Belo Horizonte MG -- Brazil}

\affiliation{$^b$ Universidade Federal de Ouro Preto -- Departamento de Matem\'{a}tica -- ICEB \\ Campus Morro do Cruzeiro, s/n, 
35.400-000, Ouro Preto MG -- Brazil}

\affiliation{$^c$ Department of Mathematical Sciences, Centre for Particle Theory, Durham University,\\
South Road, Durham DH1 3LE, United Kingdom}

\affiliation{$^d$ Departamento de F\'{\i}sica, Universidade Federal do Piau\'{\i}, Campus Ministro Petr\^{o}nio Portela, CEP 64049-550,
Teresina, PI, Brazil}

\begin{abstract}

We evaluate self-interaction effects on the quantum correlations of field modes of opposite momenta for scalar $\lambda \phi^4$ theory in a two-dimensional asymptotically flat Robertson-Walker spacetime. Such correlations are encoded both in the von-Neumann entropy defined through the reduced density matrix in one of the modes  and in the covariance expressed in terms of the expectation value of the number operators for each mode in the evolved state. The entanglement between field modes carries information about the underlying spacetime evolution.

\end{abstract}
\maketitle

\paragraph*{Introduction} - Quantum fields in non-Minkowskian spacetimes present peculiar effects as compared with quantum fields in flat spacetime \cite{BF},\cite{BD}. Of particular physical interest are dynamical spacetimes with or without horizons preventing observers from having  access to the full quantum state of a field \cite{Menicucci}. Generally, neither the particle content nor the structure of vacuum fluctuations remain covariant. The Unruh effect \cite{Unruh}, which establishes that an observer in constant linear acceleration $a$ through the Minkowski vacuum for a non-interacting quantum field will find herself immersed in a thermal bath at the temperature $T_U = \hbar a/ (2 \pi c k_B)$, is a good example. Shortly before that, Hawking \cite{Hawking} predicted that a black hole should both radiate with a temperature $T_H =  \hbar g/ (2 \pi c k_B) $ and have entropy $S=k_B c^3 A/ (4 G \hbar)$,  $g$ being the acceleration at the surface of a black hole, $G$ the Newton constant and $A$ the area of the event horizon. This fusion between statistical physics, relativity and quantum mechanics forms a landmark in the search for a quantum theory of gravity. Benchtop laboratory experiments to measure the Unruh temperature can be devised based on a connection
between quantum field theory and quantum optics: the change of coordinates in the description of the state of a scalar field between an inertial observer (free-falling into a black hole) and a noninertial observer (escaping from falling outside the event horizon) is equivalent to the transformation that affects a light beam undergoing parametric down-conversion in an optical parametric oscillator \cite{PDC}. Entanglement in particle physics has become an important tool to study CPT violation in kaon systems \cite{PP}, Bell inequalities \cite{BI}, quantum gravity \cite{QG} and conformal field theory and holography \cite{CFTH}.

Relativistic quantum information plays a key role in studying quantum cryptography, quantum teleportation, quantum computation and quantum metrology  both in inertial and noninertial frames \cite{RQIA} . More specifically, there have been a series of studies by Fuentes and collaborators pointing out  that gravity or noinertial motion may serve to enhance quantum information protocols \cite{Friis}. Quantum entanglement also serves as a tool to study experimental and theoretical cosmology. 
In the early universe, the energy content was largely dominated by highly entangled quantum field background \cite{Menicucci}. Even though experimental evidences show that primordial perturbations have undergone quantum-to-classical transition by some decoherence mechanism, some quantum correlations could in principle linger, in the case of weakly interacting fields, and encode information about the evolution of the universe \cite{Nambu},\cite{Parentani} . The theoretical framework to study such phenomena is quantum field theory in curved backgrounds \cite{BF},\cite{BD},\cite{Gustavo}-\cite{FIPC2}. As quantum technologies progress, the effects of gravity and motion on quantum correlations need to be accounted for in high precision measurements such as Unruh temperatures \cite{Aspachs}, detection of gravity waves  using Bose-Einstein condensates \cite{Sabin}, Schwarzchild parameters of the Earth \cite{SPE}, etc. The accuracy of physical parameter estimations can be improved by preparing a probe in a particular (entangled) state up to $1/N$ (Heisenberg limit), $N$ being the number of measurements (instead of $1/\sqrt{N}$ for a classical probe). The general techniques for applying quantum metrology to quantum field theory were introduced in \cite{Ahmadi1},\cite{Ahmadi2}.

The propagation of quantum fields in expanding spacetimes leads to spontaneous creation of pair of particles with
opposite momenta (alias, dynamical Casimir effect in condensed matter physics) building up nonlocal quantum correlations. In \cite{Fuentes0}, the entanglement of quantum field modes of opposite momenta was shown to contain  information about the cosmic parameters characterizing the spacetime expansion. The entanglement generated between opposite field modes during a period of spacetime expansion was calculated in terms of the cosmological parameters of a $1+1$-dimensional conformally flat Robertson-Walker (RW) model for a free scalar theory.  In the limit of small mass, this allowed for expressing the cosmological parameters in terms of the amount of the entanglement generated throughout cosmic evolution. In a similar fashion, in ref. \cite{Fuentes1} it was studied the entanglement between modes of opposite momenta of a free Dirac field. The latter showed qualitative differences as compared with the bosonic counterpart, namely that fermionic fields encodes more information about the underlying spacetime than the bosonic case. Whilst for the bosonic case the von Neumann entropy $S^B_{vN}(p ; m ) $ of an observed mode decreases monotonically from a maximum at $p=0 $,  for a Dirac field the von Neumann entropy $S^F_{vN}(p ; m) $ peaks at a certain momentum $p > 0 $ which is sensible to the rapidity of the expansion. Moreover the amount of entanglement in the maximally entangled  mode $p$ is sensitive to the total volume of expansion.

In this short letter we aim at assessing the role of  interactions in the task of evaluating cosmological parameters in the simplest possible setup: a scalar field $\lambda \phi^4$ self-interaction in an asymptotically flat $1+1$ dimensional RW expanding spacetime \cite{BF}, \cite{Fuentes0}. It is a first step towards studying a more realist interacting model.  It is generally believed that the number of particles created from the vacuum suffers either a gravitational amplification  or attenuation in the presence of field interactions depending on the spin-statistic of the particles \cite{GA},\cite{HU}. Whilst pair production from free fields 
corresponds to a mixture of positive and negative frequencies, interactions lead to multiparticle production which can enhance the amount over that of a free field. It is important firstly to evaluate interaction effects on particle production to compare with backreaction of the quantum field on the spacetime. We evaluate the entropy generated over a period of spacetime expansion to leading order in perturbation theory and estimate the interaction effect on the amount of information encoded in the entanglement of the scalar field modes.

\paragraph*{Self interaction field} - Let us consider a massive real scalar field $\phi$  in a two dimensional RW spacetime
with expansion factor $a^2(t)$. The action reads

\be 
S=\int d^2 x \sqrt{|g|} \Big( \frac{1}{2} g_{\mu \nu} \partial^\mu \phi \partial^\nu \phi - \frac{1}{2} m^2 \phi^2 - \lambda \phi^4 \Big)\, ,
\ee

\noindent where $g = \det g_{\mu\nu}$ and $\lambda$ is the coupling constant.  The dynamics of the field operator $\phi$ in the interaction picture is governed by the covariant  Klein-Gordon equation
\begin{align}
(\square + m^2) \phi  = 0, \label{EqM}
\end{align} 

\noindent $\square = (\sqrt{|g|})^{-1}\partial_\mu ( \sqrt{|g|} g^{\mu \nu} \partial_\nu)$, whereas the state evolution is controlled by the interaction Hamiltonian 
\be 
H_I = \lambda \int dx \sqrt{|g|} \phi^4 (x), 
\label{HI}
\ee
where the integration is taken over a spacelike surface of constant conformal time $\eta = \int\frac{dt}{a(t)}$. A spatially flat Robertson-Walker (RW) universe metric in $1+1$ dimensions in terms of $\eta$ is conformally equivalent to the Minkowski metric $\eta_{\mu \nu}$:
\be
ds^2 = C^2(\eta) (d\eta^2 - dx^2) .
\ee
\noindent We adopt, following \cite{Fuentes0}, 
\be 
C^2(\eta) = 1 + \epsilon (1 + \tanh (\rho \eta)), 
\ee
$\epsilon, \rho > 0$  controlling the volume and rapidity of the expansion. In curved spacetimes the existence of a unique global Killing vector $K = \partial_{\eta}$ orthogonal to all spacelike hypersurfaces  that guarantees an unambiguous separation of positive- and negative-frequency modes is not possible in general. For the scalar factor $C^2(\eta)$, we have two asymptotic regions in the distant past $C^2(\eta \rightarrow -\infty) = 1$ and in the far future $C^2(\eta \rightarrow +\infty) = 1 + 2 \epsilon$. In such asymptotic regions Poincar\'{e} invariance guarantees the existence of a unique global timelike Killing vector field $K$ and thus one can define particle states in terms of positive and negative frequency modes, as well as a vacuum state. However, in the intermediate region, particle interpretation is not defined. We assume that in the asymptotically flat regions $\lambda \sim 0$, that is the self-interaction is adiabatically switched off in the distant past and future.

Translational invariance of (\ref{EqM}) enables us to write $$\phi_k (\eta, x)= (2\pi \omega_k)^{-1/2} e^{ikx} \xi_k (\eta),$$ satisfying
\be
\partial^2_\eta \xi_k(\eta) + (k^2+C^2(\eta)m^2)\xi_k (\eta) = 0. 
\ee
The solutions of the equation above admit a simple form in the asymptotic regions \cite{BF},\cite{Fuentes0}, namely $\xi_k (\eta \rightarrow -\infty)\equiv \bar{u}_k$ and $\xi_k (\eta \rightarrow +\infty)\equiv u_k$, hereafter called ``in"- and ``out"-mode functions, respectively. In our notation, barred (unbarred) variables  represent ``in" (``out") variables. They read
\begin{align}
\bar{u}_k(\eta, x) &= \frac{1}{\sqrt{4 \pi \bar{\omega}}} e^{i (k x - \bar{\omega} \eta)},  \nonumber \\
u_k(\eta, x) &= \frac{1}{\sqrt{4 \pi \omega}} e^{i (k x - \omega \eta)},      \label{MF}
\end{align}
with $\bar{\omega} = (k^2 + m^2)^{1/2} $ and $\omega = (k^2+m^2 (1 + 2 \epsilon))^{1/2}$. Evidently we may write the in-mode functions in terms of the out-mode functions,
\be
\bar{u}_{k} (x,\eta) =  \alpha_{k} u_k (x,\eta) + \beta_k u_k^* (\eta,x), 
\ee
where the Bogolyubov coefficients $\alpha_k$ and $\beta_k$ satisfy the normalization condition $|\alpha_k|^2 - |\beta_k|^2 = 1$. In this particular case they are readily evaluated to \cite{BD}
\begin{align}
     \alpha_{k} &= \sqrt{\frac{\omega_{k}}{\overline{\omega}_{k}}}\frac{\Gamma(1 - \frac{i\overline{\omega}_{k}}{\rho})\Gamma(-\frac{i\omega_{k}}{\rho})}{\Gamma(-\frac{i\omega_{+}}{\rho})\Gamma(1 - \frac{i\omega_{+}}{\rho})}, \nonumber \\
     \beta_{k} &= \sqrt{\frac{\omega_{k}}{\overline{\omega}_{k}}}\frac{\Gamma(1 - \frac{i\overline{\omega}_{k}}{\rho})\Gamma(\frac{i\omega_{k}}{\rho})}{\Gamma(\frac{i\omega_{-}}{\rho})\Gamma(1 + \frac{i\omega_{-}}{\rho})},
\label{BC}     
\end{align}       
  where $\omega_{\pm} = \frac{1}{2}(\omega_{k} \pm \overline{\omega}_{k})$. In the interaction picture, the Bogolyubov coefficients carry information only about noninteracting contribution to the total particle creation. Moreover particle content is well defined in the asymptotic regions through their creation and annihilation operators related by
\begin{align}
a_k  &= \alpha_k \bar{a}_k + \beta_k \bar{a}^\dagger_{-k} \nonumber \\
a_k^\dagger  &= \alpha_k^* \bar{a}_k^\dagger + \beta_k^* \bar{a}_{-k}.
\end{align} 
A Fock space state $| \bar{\psi} \rangle$ in the in-region is related by the state $|{\psi} \rangle$ in the out-region by the two mode squeezing operator $\cal{S}$:
\bq
|{\psi} \rangle &=& \cal{S} |\bar{\psi}\rangle  \nonumber \\
\cal{S} &\equiv& \exp\Big(\sum_q - \zeta {a}_q^\dagger {a}_{-q}^\dagger + \zeta^*{a}_q {a}_{-q}  \Big),
\eq
$\alpha_q = \cosh\zeta$ and $\beta_q =\sinh\zeta$. For instance 
\be 
|{0}\rangle =\bigotimes_{q} \sum_{n=0}^\infty c_n^q \frac{({\bar{a}}_q^\dagger {\bar{a}}_{-q}^\dagger)^n}{n!}  |\bar{0} \rangle, 
\label{outvacuum}
\ee
which means that the (pure) vacuum in-state appears as a state with particle excitations in the out-region (and vice-versa). The Schmidt coefficients $c_n^q$ encode Bogolyubov coefficients and can be evaluated to give \cite{Fuentes0}
\be
c_n^q= \sqrt{1-\gamma_q} \Big( \frac{\beta_q}{\alpha_q} \Big)^n \, , \,\,\,\, \gamma_q =\Big| \frac{\beta_q}{\alpha_q}\Big|^2 . 
\ee
The interpretation is that particles are created as a result of cosmic expansion. 

     \paragraph*{von-Neumann entropy} - Let us compute to leading order in perturbation theory the contribution of $\lambda \phi^4$ interaction to particle production and entanglement. The modes $q$ and $-q$ define a bi-partition and we may compute the entanglement of a particular mode of the state (\ref{outvacuum}), say $q=p$, by defining the reduced density matrix $\hat{\rho}_p = {\mbox{tr}}_{-p} ( |\bar{0}\rangle_{p,-p} \,\, {}_{p,-p}\langle \bar{0}| ) $ and calculating the von-Neumann entropy 
\be 
S_{vN} = - {\mbox{tr}}( \hat{\rho}_p \log_2 \hat{\rho}_p ).
\label{SVN}
\ee
In the free case \cite{Fuentes0}, 
\be 
S_{vN} = \log_2 \Big( \frac{\gamma_p^{\gamma_p/(\gamma_p-1)}}{1-\gamma_p}\Big) ,
\ee 
$0\le S_{vN} < 1$, which can be inverted $\gamma_p = \gamma_p (S_{vN})$ to write a simple expression for the cosmological parameters
$\epsilon = \epsilon (\gamma_p)$ and $\rho = \rho (\gamma_p)$ in the limit where 
$m<<2 \rho \epsilon^{-1/2}$, showing that information about spacetime is stored in entanglement of very light particles produced over cosmic expansion.  

\begin{figure}[htp]
\begin{center}
 \includegraphics[scale=0.8]{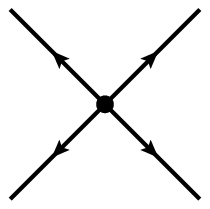} \quad\qquad \includegraphics[scale=0.8]{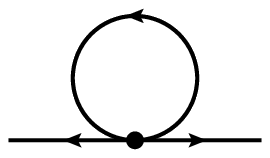}
\caption{Quartet creation (left) and pair creation (right) by self-interaction.}\label{Diagram}
\end{center}
\end{figure}

  In the interaction picture, such an information about particle production is enclosed in the state vector. Let us write the in-vacuum state to $O(\lambda)$ as
\bq
|\bar{\psi}\rangle &=& N \Big(|\bar{0}\rangle + \frac{1}{2!}\int_{k_1,k_2} \Gamma_{1,2} |{\overline{k_1 k_2}}\rangle\nonumber \\
&+& \frac{1}{4!}\int_{k_1,k_2,k_3,k_4} \Gamma_{1,2,3,4} |{\overline{k_1 k_2 k_3 k_4}}\rangle \Big),
\label{VC}
\eq 
where $\int_k \equiv \int dk$ and
\bq
\Gamma_{1,2} &=& \langle \overline{k_1 k_2} |-i \int_{-\infty}^{+\infty} H_I \, d\eta |\bar{0}\rangle,\nonumber \\
\Gamma_{1,2,3,4} &=& \langle \overline{k_1 k_2 k_3 k_4}|-i \int_{-\infty}^{+\infty} H_I \, d\eta |\bar{0}\rangle ,
\label{Gammas}
\eq 
with $H_I$ given by (\ref{HI}) and depicted in figure \ref{Diagram}.
     
Applying the squeezing operator ${\cal{S}}$ on both sides of (\ref{VC}) and using that
\bq
&&{\cal{S}} |\overline{k_1 .. k_n}\rangle = {\cal{S}} \bar{a}^\dagger_{k_1} {\cal{S}^\dagger}{\cal{S}}\ldots \bar{a}^\dagger_{k_n} {\cal{S}^\dagger}{\cal{S}}|\bar{0} \rangle =\nonumber \\ 
&&a^\dagger_{k_1} \ldots a^\dagger_{k_n} |0 \rangle  = 
(\alpha_{k_1}^* \bar{a}_{k_1}^\dagger +\beta_{k_1}^* \bar{a}_{-k_1}) \nonumber \\ && \ldots(\alpha_{k_n}^* \bar{a}_{k_n}^\dagger +\beta_{k_n}^* \bar{a}_{-k_n}) \bigotimes_q \sum_n c_n^q |\bar{n}_q \bar{n}_{-q} \rangle,
\eq 
and  $\bar{\Gamma}_{1,\ldots,n} = \Gamma_{1,\ldots,n}$ enable us to construct the out density matrix $\hat{\rho}_{p, -p}$ after picking up a particular mode partition $q = p$  in $\cal{S}|\bar{\psi}\rangle \langle \bar{\psi}| {\cal{S}}^\dagger$ allowed by the simple mixing properties of the Bogolyubov transformations. The piece that corresponds to pair creation in (\ref{Gammas}) is divergent and needs regularization ($\Lambda$) and renormalization. It is easy to see  that $\Gamma_{1,2}^\Lambda = - 12 i \lambda \langle \phi^2 \rangle^\Lambda \int d^2x \,\, \sqrt{|g|} u_{k_1}^* u_{k_2}^*$  \cite{BF}. Adding a counterterm $\frac{1}{2} \delta m^2 \phi^2$ in the Lagrangian and choosing $\delta m^2 = - 12\lambda \langle \phi^2 \rangle^\Lambda $  renormalizes the pair creation to zero. Therefore only the quartet creation process will contribute. The non-vanishing terms contributing to the reduced density matrix read
\begin{align}
&\hat{\rho}_p = \sum_m \langle \bar{m}_{-p}| \hat{\rho}_{p, -p} |\bar{m}_{-p} \rangle = \sum_n |c_n^p|^2\times \nonumber \\
& \Big( 1 + \lambda (10 n^2 + 14 n + 6)  \mbox{Re} [(\alpha_p^{*})^2(\beta_p^{*})^2 B(p)]\Big) |\bar{n}_p\rangle \langle \bar{n}_p|,                 \label{RDM}
\end{align} 
where we have defined $$\Gamma_{1,2,3,4}= \lambda \delta (k_1+k_2+k_3+k_4) B(k_1,k_2,k_3,k_4)$$ and  $B(p)=B(k_1,\ldots,k_4)$ for $k_i \rightarrow \pm p$. In two dimensions $B \neq 0$ as the self interaction destroys conformal invariance. The first term in (\ref{RDM}) is just the free field contribution obtained in \cite{Fuentes0} while the second term can be evaluated using the asymptotic mode functions in (\ref{MF}) to give 
\begin{align}
B(p) &= \frac{i}{4 \bar{\omega}}\int_{-\infty}^{+\infty} d\eta \, C^2(\eta) e^{-4i\bar{\omega} \eta} \nonumber \\
  &= \frac{i\pi(1+\epsilon)\delta(4\bar{\omega})}{2 \bar{\omega}} + \frac{i\epsilon}{4 \bar{\omega}}f(\bar{\omega}).
\label{INTB}
\end{align}
where $\bar{\omega} = \sqrt{p^2+m^2}$ and 
\begin{align}
f(\bar{\omega}) &= \int_{-\infty}^{+\infty} d\eta\tanh(\rho\eta)e^{-4i\bar{\omega} \eta}, \nonumber \\
   &= \frac{1}{\rho}\int_{-\infty}^{+\infty} d\tau\tanh(\tau)e^{-i\Omega\tau}, \quad \Omega \equiv \frac{4\bar{\omega}}{\rho}. \nonumber
\end{align}
  The first term proportional to the delta function gives no contribution to the reduced density matrix (it amounts to a shift in the scale factor) \cite{BF}. The integral expressed by $f(\bar{\omega})$ is a little delicate due to convergence at $\tau \rightarrow \pm\infty$. So, consider instead a regularized version 
$\Omega \rightarrow  \Omega - i\varepsilon\kappa(\tau)$, with $\varepsilon > 0$, $\kappa(\tau) = \frac{\tau}{|\tau|}$. The integral $f_{\varepsilon}(\Omega)$ now converges for $\tau \rightarrow \pm\infty$ and after an integration by parts we obtain
\begin{align}
 \lim_{\varepsilon \to 0} f_{\varepsilon}(\Omega) = -\frac{i}{\rho\Omega}\int_{-\infty}^{+\infty} d\tau\frac{e^{-i\Omega\tau}}{\cosh^2(\tau)}. \nonumber
\end{align}
  This integral can be evaluated in the complex plane by replacing the real variable $\tau$ by the complex variable $z$ and integrating around a closed counterclockwise rectangular contour $\Sigma$ with vertices at $-\Lambda$, $\Lambda$, $\Lambda + i\pi$, $-\Lambda + i\pi$. In the upper horizontal  path we use that $\cosh^2(\tau) = \cosh^2(\tau + i\pi)$. The two vertical paths parallel to the imaginary axis vanish exponentially as $\Lambda \rightarrow \infty$. 
Thus
\begin{align}
 \lim_{\varepsilon \to 0} f_{\varepsilon}(\Omega)  = \frac{i}{\rho\Omega}\frac{1}{ e^{\Omega\pi} - 1}\oint_\Sigma dz \frac{e^{-i\Omega z}}{\cosh^2(z)},
\end{align}
whose integrand has a pole at $z = \frac{i\pi}{2}$. Using Cauchy's integral formula we find
\begin{align}\label{INTB2}
B(p) =  \frac{\epsilon}{4\bar{\omega}\rho}\frac{\pi}{\sinh(\frac{2\pi\overline{\omega}}{\rho})}.
\end{align}
    This term expresses a thermal like profile of the vacuum generated by the interaction over the spacetime evolution. 
    
    The quantum entanglement between the field modes $-p$ and $p$ for the interacting theory can be quantified by the von-Neumann entropy (\ref{SVN}) using the reduced density matrix given by (\ref{RDM}) and the Bogolyubov coefficients (\ref{BC}). We plot some graphs to illustrate its behaviour as a function of the field mode and mass in comparison with the free case.

\begin{figure}[htp]
\begin{center}
 \includegraphics[scale=0.21]{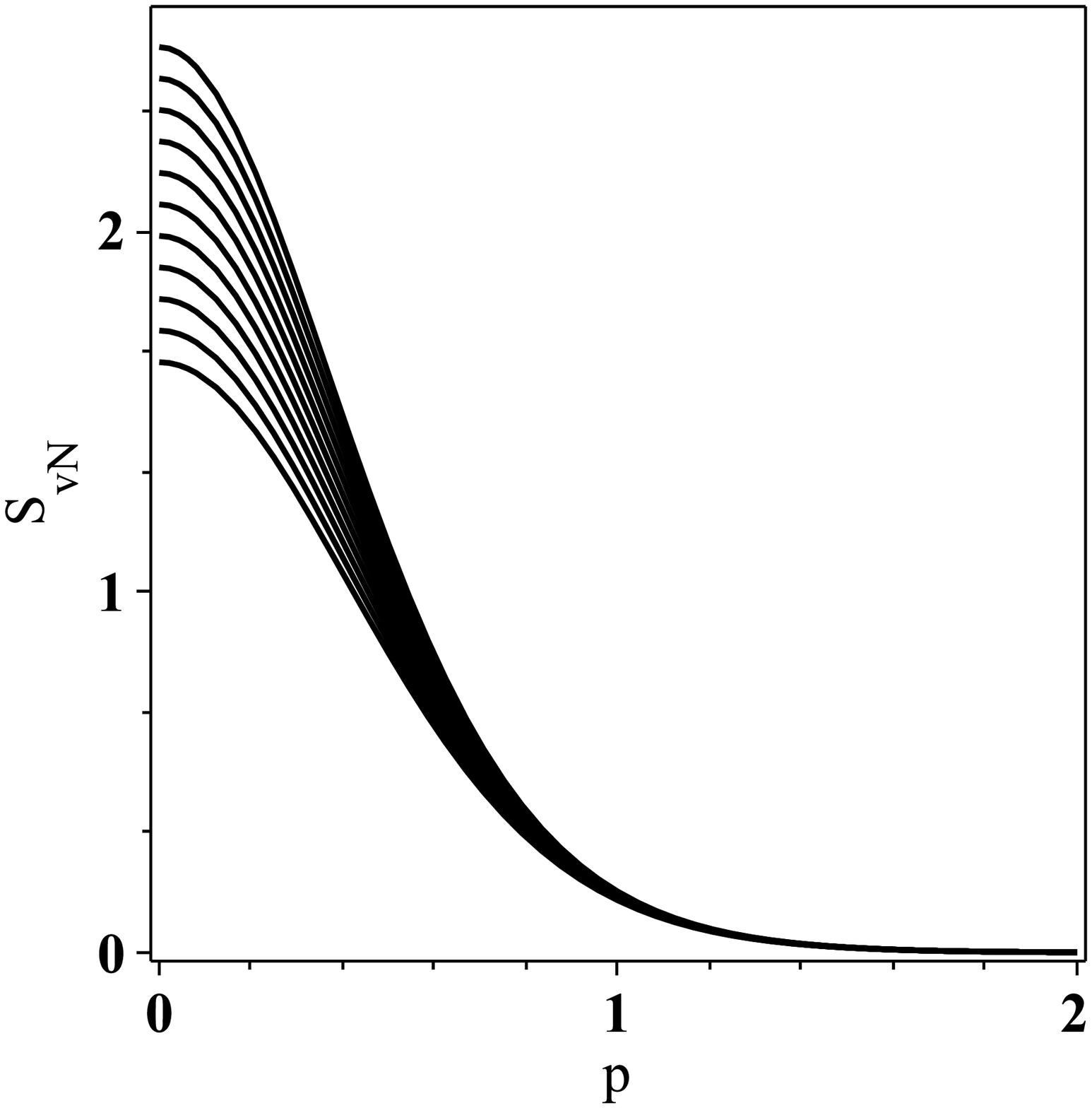} \quad \includegraphics[scale=0.21]{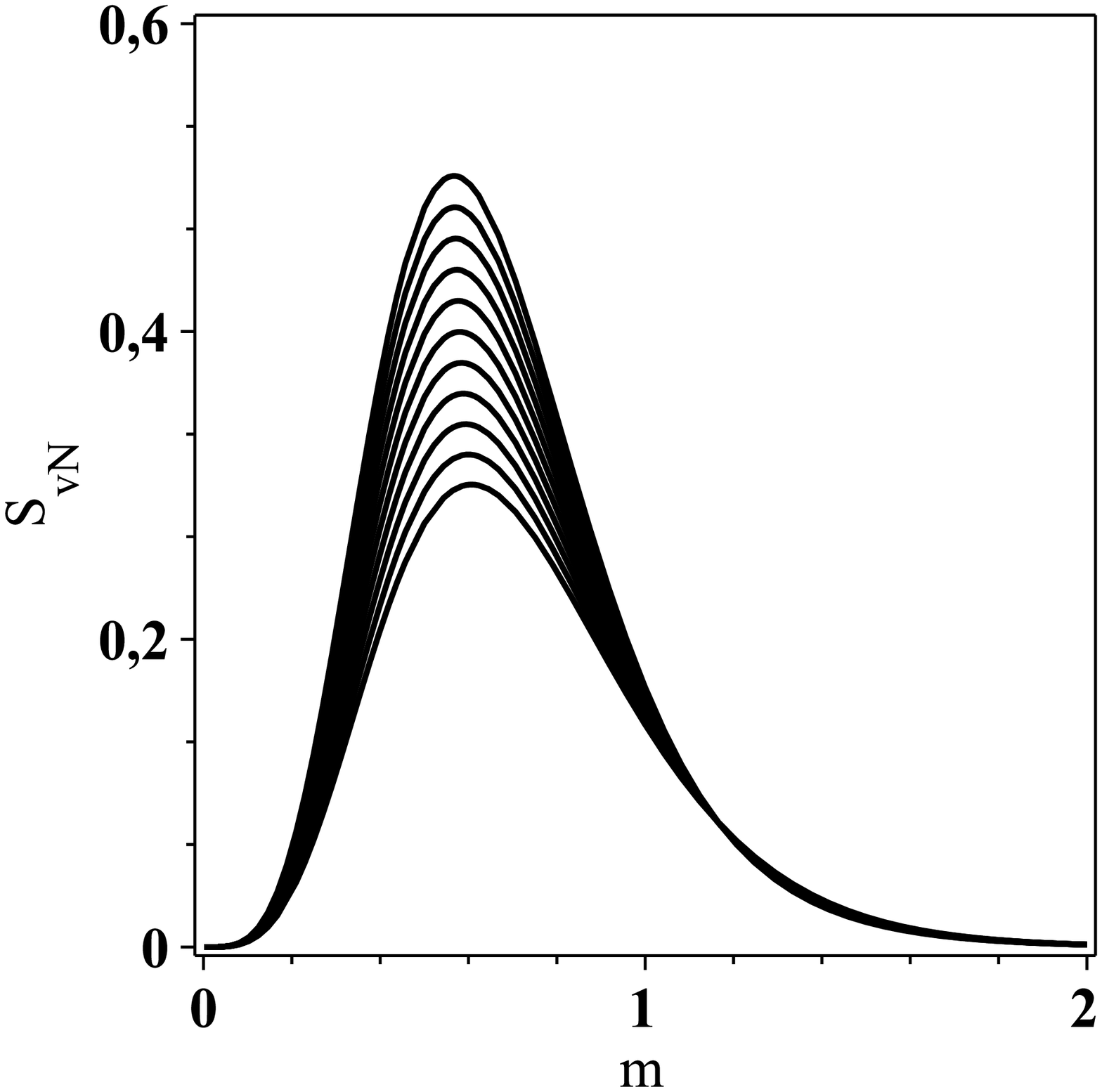}
\caption{Numerical evaluation of entanglement entropy as function of the $p$ (left) with $\rho = $ $m=$ $\epsilon = 1$ and as function of the $m$ (right) with $\rho = p = \epsilon = 1$, both for different coupling constants $0 \leq  \lambda < 1$. Higher peaks correspond to stronger couplings.}\label{fig4}
\end{center}
\end{figure}

  
   The spectral behaviour of the entanglement entropy $S_{vN}$ suggests that the effects of such coupling will be important for small values of $p$. This means that the self interaction amplify the low frequency modes of the scalar field producing an enhancement in the quantum correlation between the particle pairs created by spacetime expansion.

   \paragraph*{Covariance measures} - Another way to quantify the quantum correlation between the number states of modes $p$ and $-p$ is examine the covariance of the number operators defined as 
\be 
   \langle [\Delta(\hat{n}_{p} - \hat{n}_{-p})]^2\rangle = \langle (\Delta \hat{n}_{p})^2\rangle + 
\langle (\Delta \hat{n}_{-p})^2\rangle - 2\mathrm{Cov}(\hat{n}_{p},\hat{n}_{-p}), \label{Vdif} \nonumber
\ee
in which
\be 
   \mathrm{Cov}(\hat{n}_{p},\hat{n}_{-p}) = \langle \hat{n}_{p}\hat{n}_{-p}\rangle - 
   \langle \hat{n}_{p}\rangle\langle \hat{n}_{-p}\rangle .
\ee
    For states containing no intermode correlations the covariance vanishes. for the  two-mode squeezed vacuum state (\ref{outvacuum}), we obtain $\mathrm{Cov}(\hat{n}_{p},\hat{n}_{-p}) = |\alpha_{p}|^2|\beta_{p}|^2$, the maximal value indicating strong intermode correlations. 
    
    In the equations above $\hat{n}$ is the out number operator. Using the Bogolyubov transformations we may write
\begin{align}
   \hat{n}_{p} &= a_{p}^{\dagger}a_{p} 
   = |\alpha_{p}|^2 \bar{a}_{p}^{\dagger} \bar{a}_{p} + \alpha_{p}^{*}\beta_{p}\bar{a}_{p}^{\dagger}\bar{a}_{-p}^{\dagger}\nonumber \\
   &+ \alpha_{p}\beta_{p}^{*}\bar{a}_{-p} \bar{a}_{p} + |\beta_{p}|^2 \bar{a}_{-p} \bar{a}_{-p}^{\dagger},
\end{align}  
and a similar expression for $\hat{n}_{-p}$. For instance, the expectation values in (\ref{Vdif}) are taken with respect to the full in-state (\ref{VC}) in the interaction picture. After some straightforward algebra we get
   \be 
   \mathrm{Cov}^{(\lambda)}(\hat{n}_{p}, \hat{n}_{-p}) = |\alpha_{p}|^2|\beta_{p}|^2   
   + \frac{\lambda}{12}\mathrm{Re}[(\alpha_{p}^{*}\beta_{p})^2 B(p)]. \nonumber
   \ee 
  The first term in the equation above refers to quantum entanglement between the modes of the free field, while the second reflects the effect of the interaction which produces a gravitational enhancement on the correlations of modes as we see in figure \ref{fig4}. Notice that modes of a massless quantum field get entangled, due to the fact that in two dimensions the self interaction destroys the conformal invariance of the theory.

\begin{figure}[htp]
\begin{center}
 \includegraphics[scale=0.21]{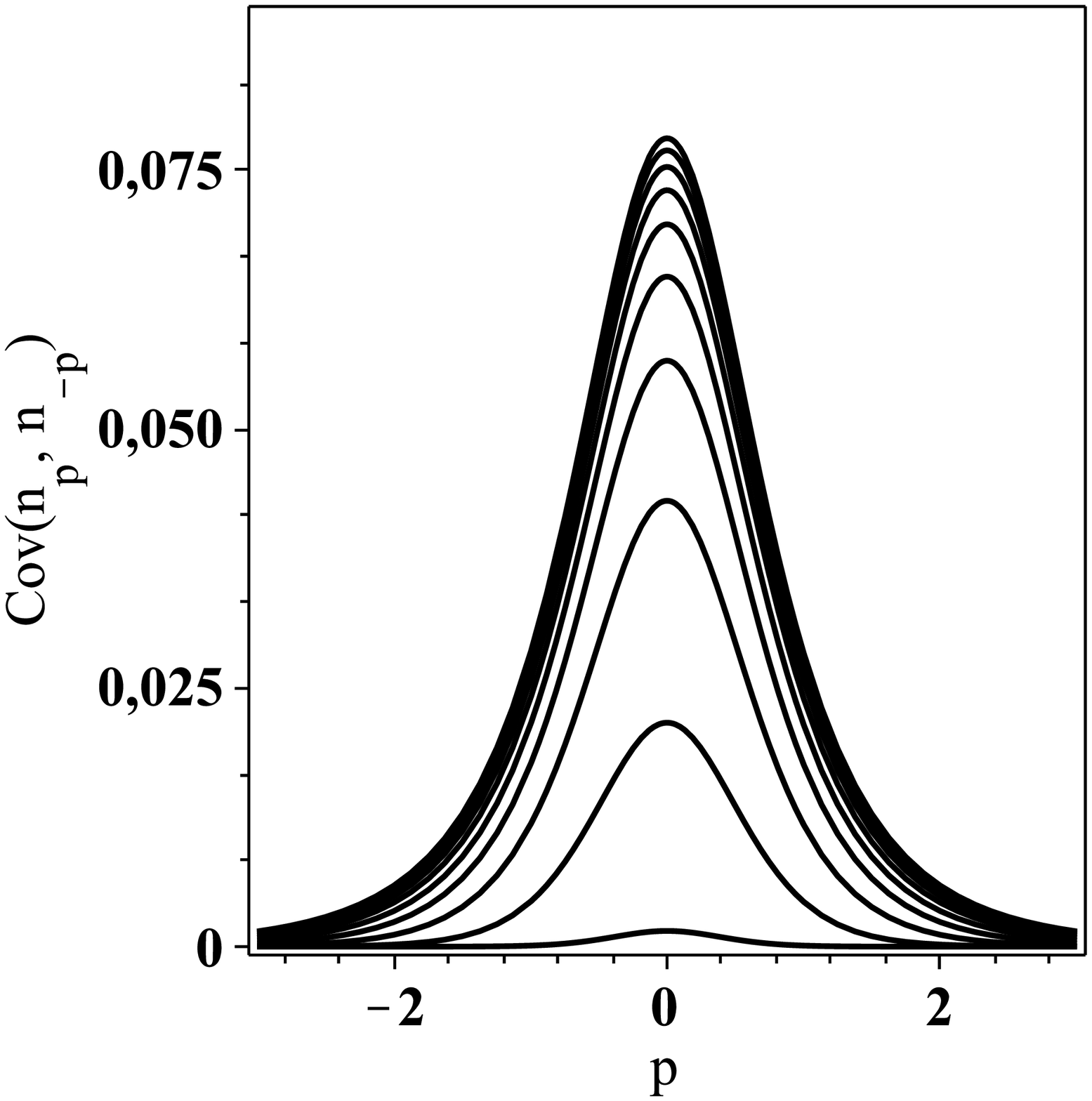} \quad \includegraphics[scale=0.21]{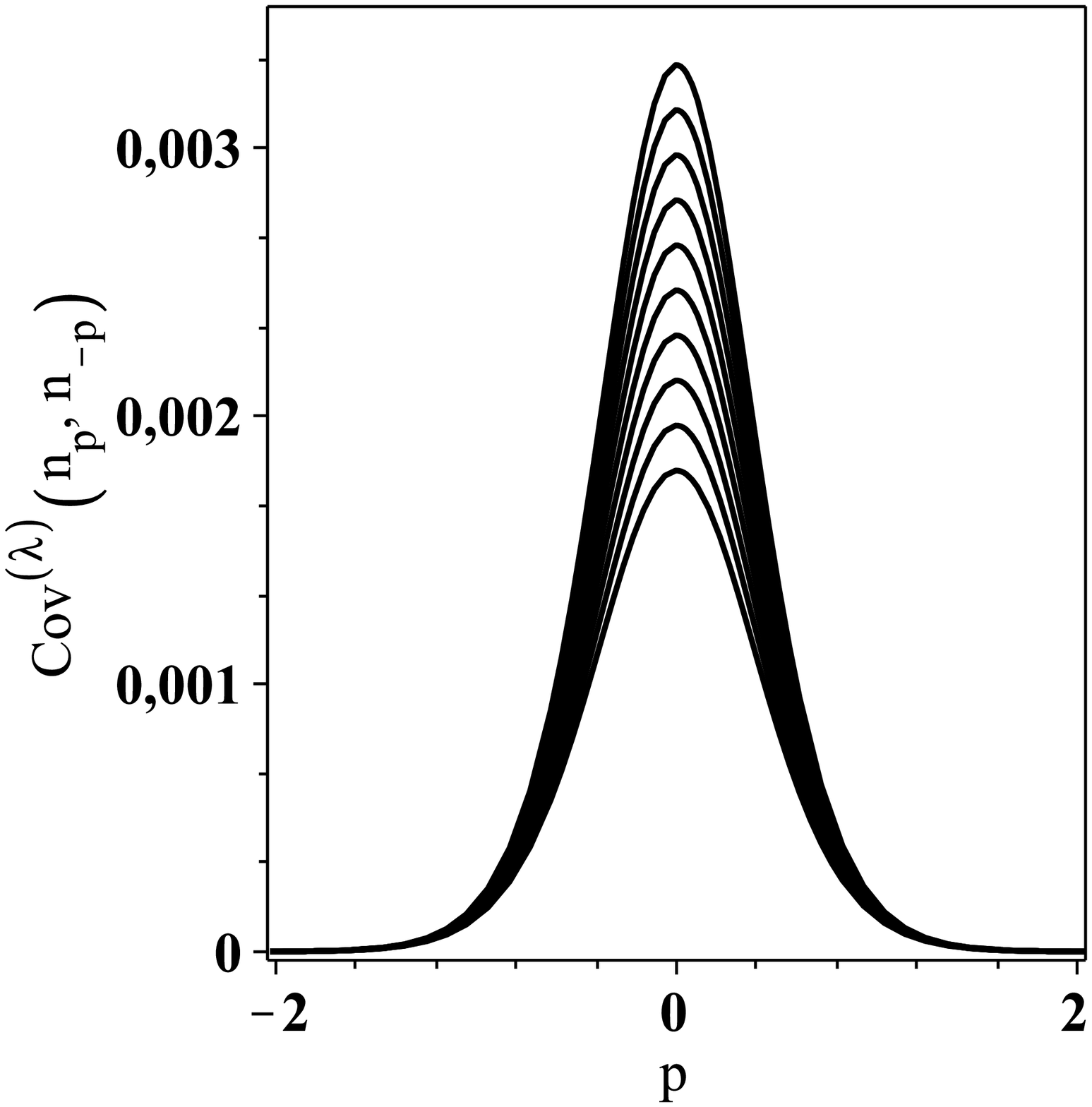}
\caption{$\mathrm{Cov}^{(\lambda)}(\hat{n}_{p}, \hat{n}_{-p})$ as function of the $p$ for different expansion parameter 
$\rho = 1 \ldots 10$ (left) with $\lambda = 0$, $m=1$ and $\epsilon = 1$ and for different coupling constants $0 \leq  \lambda < 1$ (right), and  $m = \rho = \epsilon = 1$.}\label{fig4}
\end{center}
\end{figure}

\par The intermode correlation increases for small values of $p$ when the cosmological parameter $\rho$ grows saturating 
at $\rho \approx 10$ which is also encoded in the von-Neumann entropy. This enhancement in the correlation between modes of opposite momenta  is combined effect of cosmic expansion through the Bogolyubov transformations and the interaction generating a nontrivial coupling between the modes of the scalar field. 

 \paragraph*{Discussion and conclusions} - We studied the effect of the self interaction on the entanglement between the modes of a scalar field due to spacetime expansion. We found that the quantum correlation intermode undergoes an enhancement generated by interaction. The main point of the discussion is that this enhancement favoring the encoding of information about the underlying spacetime in the entanglement intermode. It would be intersting to pursuit this dynamics for fermionic fields which are known to be more effective for extracting information about the spacetime evolution. In particular, it would be intersting to build as interaction model for fermions in which we can treat the auxiliary field as an enviroment to assess the deleterious effects of upon field mode correlation.

                            $$\ast \ast \ast$$   

 H. A. and M. S. acknowledge financial support from CNPq (Brazil). P. M. acknowledges financial support from STFC under the consolidated Grants ST/J000426/1 and ST/L000407/1. His research is also supported in part by the Marie Curie network GATIS (gatis.desy.eu) of the European Union's Seventh Framework Programme FP7/2007-2013/ under REA Grant Agreement No 317089.

\end{document}